\documentclass[conference]{IEEEtran} 
\pdfoutput=1

\usepackage{graphicx,cite}
\usepackage[cmex10]{amsmath}
\usepackage{url}
\usepackage{subfigure}

\usepackage{amssymb}
\usepackage{array}
\usepackage{fixltx2e}
\usepackage{amsfonts}
\usepackage{lineno}

\begin{document}

\title{\huge Enabling Relaying Over Heterogeneous Backhauls\\ in the Uplink of Wireless Femtocell Networks
}

\author{
\IEEEauthorblockN{Sumudu Samarakoon\IEEEauthorrefmark{1}, Mehdi Bennis\IEEEauthorrefmark{1}, Walid Saad\IEEEauthorrefmark{2} and Matti Latva-aho\IEEEauthorrefmark{1} \\}
\IEEEauthorblockA{\small\IEEEauthorrefmark{1}Centre for Wireless Communications, University of Oulu, Finland, \\ email: \url{spsamare@live.com}, \url{bennis@ee.oulu.fi}, \url{matti.latva-aho@ee.oulu.fi} \\
\IEEEauthorrefmark{2}Electrical and Computer Engineering Department, University of Miami, Coral Gables, FL, USA, email: \url{walid@miami.edu}}
}

\maketitle
\nopagebreak[4]
\begin{abstract}
In this paper, we develop novel two-tier interference management strategies that enable macrocell users~(MUEs) to improve their performance, with the help of open-access femtocells. To this end, we propose a rate-splitting technique using which the MUEs optimize their uplink transmissions by dividing their signals into two types: a coarse message that is intended for direct transmission to the macrocell base station and a fine message that is decoded by a neighboring femtocell and subsequently relayed over a heterogeneous (wireless/wired) backhaul. For deploying the proposed technique, we formulate a non-cooperative game between the MUEs in which each MUE can decide on its relaying femtocell while maximizing a utility function that captures both the achieved throughput and the expected backhaul delay. Simulation results show that the proposed approach yields up to $125\%$ rate improvement and up to 2 times delay reduction with wired backhaul and, $150\%$ rate improvement and up to 10 times delay reduction with wireless backhaul, relative to classical interference management approaches, with no cross-tier cooperation.
\end{abstract}

\section{Introduction}\label{intro}
The proliferation of wireless services with stringent quality-of-service requirement is driving network operators to search for new solutions for wireless users and their serving stations closer to one another so as to enhance the coverage and capacity of next-generation wireless systems. In this respect, the deployment of small cells overlaid on existing cellular networks and serviced by low-cost, low-power, femtocell base stations~(FBS) has emerged as a promising technique for improving the indoor wireless coverage, offloading data from the macro-cellular network, and enhancing the capacity of wireless systems~\cite{pap:vikram08}. While femtocells are poised to boost the overall network spectral efficiency, their ad hoc deployment coupled with their operation over the licensed spectrum raises many important challenges such as interference management, network planning, and resource allocation.

Existing literature has studied a number of important problems in femtocell networks ranging from cross-tier and co-tier interference management, coverage hole minimization, macrocell traffic offloading, mobility management, and security, among others~\cite{pap:jeff12}. One promising technique for enhancing the co-existence between macro-cell and femtocell networks is by allowing a certain level of cooperation between these two tiers. In particular, recent studies have shown that allowing the femtocell base stations to relay part of the macro-cell users' messages can lead to an improved data rates and a more efficient network co-existence~\cite{onln:eric11}. However, allowing such cooperative techniques requires an efficient backhaul that connects the femtocell and macro-cell tiers.  As discussed in~\cite{pap:ping11,pap:johnp10,pap:ivana11}, the nature and properties of this backhaul will strongly impact the overall network performance as well as the potential performance gains from cooperation. For example, in \cite{pap:ping11}, the authors derive tractable cooperation costs that take into account two types of backhauls: wired and over-the-air (OTA). In \cite{pap:johnp10}, the authors study the impact of a TDMA-based backhaul on resource allocation in cognitive femtocell networks. More recently, the impact of the backhaul on resource allocation is studied in~\cite{pap:ivana11} in which the authors show that a wireless backhaul can be more suitable, under certain network conditions.

Clearly, while cooperation between the femtocell and macrocell tiers is expected to yield important performance improvements in next-generation small cell networks, these improvements are essentially limited by the choice of an appropriate backhaul. In fact, the heterogeneous and unreliable nature of the femtocell backhaul leads to a fundamental question: should the femtocells use an over-the-air (in-band) backhaul which requires significant spectrum resources but can guarantee reasonable delays or should they use a wired backhaul which does not require any spectrum resources but could lead to significant traffic delays? The answer to this question is particularly important in order to enable advanced techniques such as femtocell relaying.

The main contribution of this paper is to propose a novel approach for interference management which leverages cooperation between the macro-cell and femtocell tiers while jointly optimizing the choice of an appropriate backhaul supporting this cooperation. In the proposed approach, macrocell users can seek the help of neighboring open-access femtocells in order to improve their uplink data rate while taking into account the constraints introduced by an underlying heterogeneous backhaul. To this end, we propose an approach based on rate-splitting in the uplink in which the macrocell user's~(MUE)  message is split into two parts: {\it (i) a coarse message} which can only be decoded only the macrocell base station~(MBS) and {\it (ii) a fine message} which is broadcasted by the MUEs and decoded by neighboring femtocell base station~(FBS). To benefit from rate-splitting, the MUEs must appropriately select the best FBSs for relaying their signals over the backhauls and optimally split their rates between MBS and FBS. We show that these choices lead to a non-cooperative game between the MUEs in which each MUE needs to select its preferred relaying femtocell along with the associated power allocation so as to maximize its utility function which captures the tradeoff between the achieved data rate (due to relaying) and the expected transmission delay (due to the backhaul constraint). To solve this game, we propose a best response-based algorithm using which the MUEs can reach the equilibrium of the game. Simulation results show that the proposed approach outperforms classical interference mitigation approach  (with no macro-femtocell coordination) for both wired and wireless backhauling scenarios, with up to $125\%$ and $150\%$ improvements in rates and up to $5$ and $10$ times reduction in transmission delays for wired and wireless backhauls, respectively.

The rest of the paper is organized as follows. The system and backhaul models in Section~\ref{sec:sys_mod}. Section~\ref{sec:prop_mthd} presents the formulation of the proposed cooperative relaying technique over heterogeneous backhauls. The proposed game theoretical approach for relaying is discussed in Section~\ref{sec:game}. Simulation results are presented and analyzed in Section~\ref{sec:results}. Finally, conclusions are drawn in Section \ref{sec:conclu}.
\section{System Model} \label{sec:sys_mod}
\subsection{Network Model}
Consider the uplink transmission of an orthogonal frequency division multiple access and time division multiple access (OFDMA/TDMA) two-tier wireless network with a single macrocell and $F$ FBSs.
The MBS is located at the center of a cell with radius  $R_m$. The MBS serves a set of MUEs denoted by $\mathcal{M} = \{1,\ldots,M\}$ in a TDMA manner. Let  $\mathcal{N} = \{1,\ldots,N\}$ denote the set of sub-channels. Likewise, femtocell users (FUEs) communicate in the uplink with their respective FBSs in the set $\mathcal{F} = \{1,\ldots,F\}$, with each FBS $f \in \mathcal{F}$ having a radius $R_f$. We assume a static sub-channel allocation for FUEs such that each user that is serviced by FBS $f$ is assigned a single sub-channel from $\mathcal{N}$. Thus, for a sub-channel $n$ at a given time slot, there will be
a set of FUEs consisting of a single active FUE from each femtocell - denoted by $\mathcal{K}^n$. We let $|h_{ji}^{n}|^2$ denote the channel gain between transmitter $j$ and receiver $i$ on sub-channel $n$. We denote the MBS with index $0$. The transmission power of the $j$-th transmitter over $n$-th sub-channel is $P_{j}^{n}$, and the variance of the complex Gaussian thermal noise at the receiver is denoted by $N_0$.

In the classical macro-femtocell deployment scenario, there is no cooperation/coordination between the macrocell and femtocell tiers, and  hence the achievable rates of MUE $m\in\mathcal{M}$ and FUE $k\in\mathcal{K}^n$ serviced by MBS and FBS $f$ respectively are:
\begin{align} \label{eqn:noncop_mbsrate}
	&[R_{m}^{n}]_{\rm CLA} = \log_2 \biggl( 1 + \frac{|h_{m0}^{n}|^2P_m^n}{N_0 +  \sum_{\forall j\in\mathcal{K}^n}|h_{j0}^{n}|^2P_{j}^n } \biggr),\\
	&[R_{k}^n]_{\rm CLA} = \min\Biggl\{\log_2 \biggl( 1 + \frac{|h_{kf}^{n}|^2 P_{k}^n}{N_0 + I_k^n } \biggr), C_f\Biggr\},
\end{align}
where $I_k^n = |h_{mf}^{n}|^2P_m^n + \sum_{\substack{\forall j\in\mathcal{K}^n\\j\neq k\\~}}|h_{jf}^{n}|^2P_{j}^n$ is the aggregate interference experienced by the $k$-th FUE and $C_f$ is the  fixed backhaul capacity between  the $f$-th FBS and the MBS. The subscript ``CLA'' is used to denote the rates that are calculated for classical macro-femto deployment scenario.

\subsection{Backhaul Model}
In order to improve their rates in (\ref{eqn:noncop_mbsrate}), the FUEs and MUEs can cooperate and coordinate their transmissions. However, one of the key challenges for deploying cooperation in femtocell networks is to design an adequate backhaul  that can lead to an efficient communication between the macro and  femtocell tiers. In fact, the reliability of the backhaul connection between FBSs and MBSs is instrumental in the optimal deployment of  heterogeneous networks, hence requiring designs that jointly account for access and backhaul links. In practice, existing research \cite{pap:johnp10,pap:osvaldo10,pap:ping11,pap:ivana11} suggests two possible types of backhaul networks: wired and wireless. On the one hand, a wired backhaul can provide a reliable platform for communication which does not require any spectral resources, but it often leads to increased delays due to the presence of traffic from various sources. On the other hand, a wireless backhaul provides congestion-free communication but it requires additional spectrum resources and can lead to an increased interference in the network. To this end, before delving into the details of the proposed cooperative approach, we present the considered models for the two backhaul types.

\subsubsection{Wired backhaul}
We consider that the packet generation process at the femtocells follows a Poisson distribution, and, thus, we model the entire backhaul of the system as an M/D/$1$ queue~\cite{book:dim92}. Let $C_f$ be the capacity of the $f$-th FBS- MBS link and, thus, the total wired backhaul capacity $\overline{C}$ could be given by:
\begin{equation}\label{eqn:totwiredcapacity}
	\textstyle  \sum_{\forall f\in\mathcal{F}}C_f \leq \overline{C} .
\end{equation}
\subsubsection{Wireless backhaul}
In this scenario, we account for the  increased interference over FBSs-MBS  backhaul links due to femtocell transmissions over the backhaul with power $P_{f}^n$. Here, the rate of FBS $f$ is given by:
\begin{equation}
	\label{eqn:rateBackhaul}
	R_{f0}^n = \log_2 \Biggl( 1 + \frac{|h_{f0}^{n}|^2P_{f}^n}{N_0 +  \sum_{ \forall l\in\mathcal{F},l\neq f}|h_{l0}^{n}|^2P_{l}^n } \Biggr).
\end{equation}

\section{Cooperative Relaying between the Macrocell and Femtocell Tiers}\label{sec:prop_mthd}
\begin{figure}[!t]
\centering
\includegraphics[width=8cm]{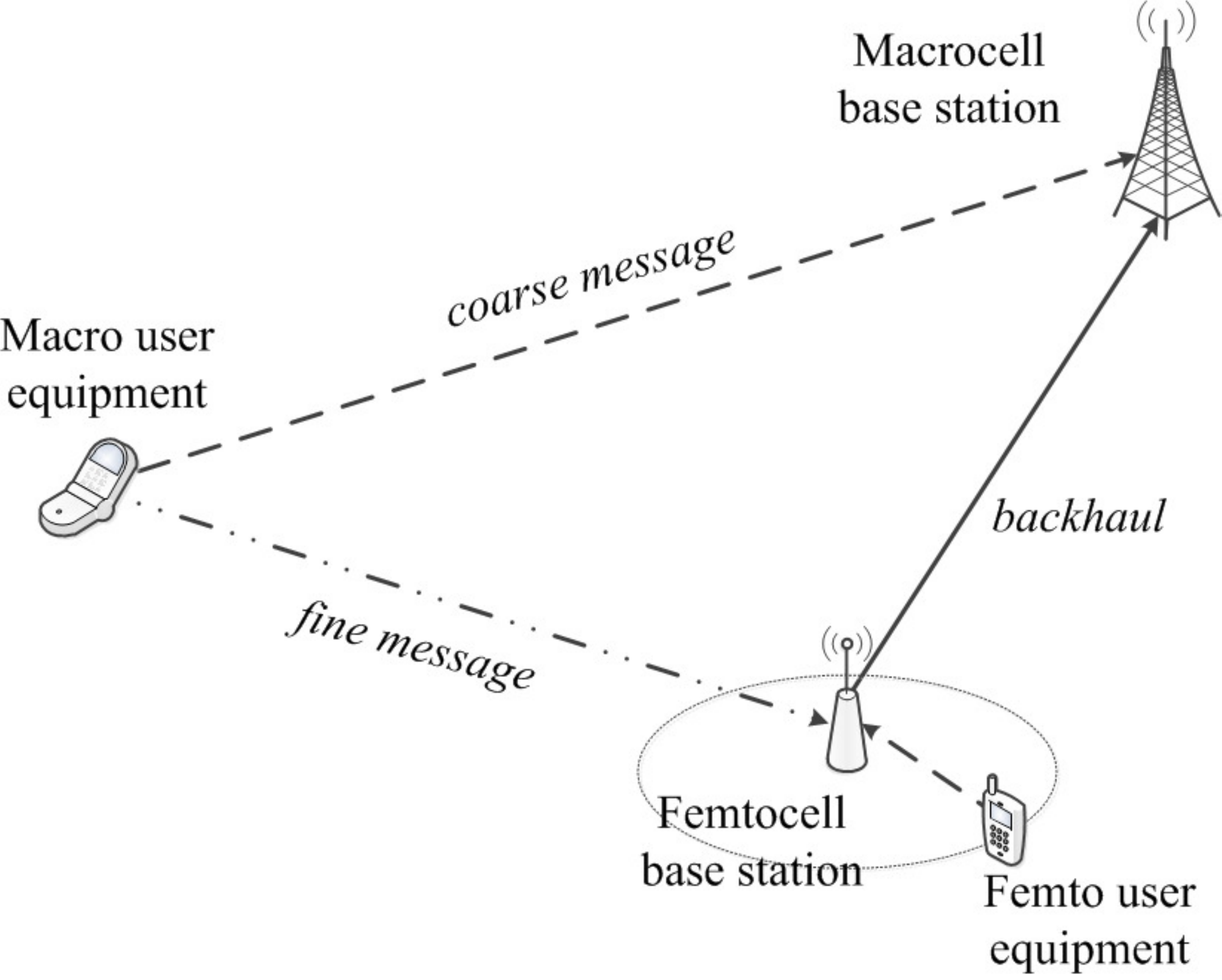}
\caption{Illustration of the proposed relaying approach in which the MUEs can use rate splitting techniques in their uplink transmission.}
\label{fig:model}
\end{figure}

To enable an efficient co-existence between the two network tiers, we propose a cooperative approach using which existing femtocells can assist  nearby MUEs in order to improve the overall data rates via the concept of rate splitting \cite{pap:randa11,pap:osvaldo10,pap:osvaldo09}. In this context, each MUE $m \in  \mathcal{M}$ builds a {\it coarse} $X_{m,C}^{n}$ and a {\it fine} message $X_{m,F}^{n}$ (direct signal and relayed signal, respectively) for each of its transmitted signals as illustrated in Fig.~\ref{fig:model}. With these two messages, the source MUE superimposes two codewords and, thus, the transmission rates associated with these messages are such that the FBSs can reliably decode the fine message while the MBS decodes the coarse message. Mathematically, this  can be expressed as follows:
\begin{equation} \label{eqn:coarseNfine}
	 X_{m}^{n}=X_{m,C}^{n}+X_{m,F}^{n}.
\end{equation}
Moreover, the transmission power allocations of the MUE's coarse signal to the MBS and the fine signal for FBSs  are $P_{m,C}^{n}=(1-\theta)P_m^n$ and $P_{m,F}^{n}=\theta P_m^n$, respectively, with $0\leq\theta<1$.

In this proposed scheme, a femtocell with both a good channel gain from a neighboring MUE and a high backhaul capacity can assist the MUE by first decoding, and subsequently relaying its fine message to the MBS over the backhaul. Subsequently, the overall MUEs' transmission rates are improved. Here, we consider that, upon relaying MUE's fine messages,  the FBSs can simultaneously service their own FUEs using successive interference cancellation (SIC)~\cite{book:tse05}.

\subsection{Wireless over-the-air (OTA) backhaul}
We assume a half duplex decode and forward (DF) uplink transmission in which both MUEs and FUEs transmit during the first time slot and FBSs relay both signals (over the backhaul) during the second time slot. The uplink rate   of MUE $m\in \mathcal{M}$ when transmitting its coarse message to its serving MBS over sub-channel $n \in \mathcal{N}$ is given by:
\begin{align}
\label{eqn:rateMacroPriv}
	[R_{m,C}^{n}]_{\rm OTA} &= \log_2 \Biggl( 1 + \frac{|h_{m0}^{n}|^2(1-\theta)P_m^n}{N_0 + \sum_{\forall j\in\mathcal{K}^n}|h_{j0}^{n}|^2P_{j}^n } \Biggr),
\end{align}
where $(1-\theta)P_m^n$ is the MUE's transmission power allocated for the coarse message.
Similarly, the rate of MUE $m$ when  transmitting its fine message to  FBS $f \in \mathcal{F}$ over sub-channel $n$ is given by:
\begin{equation}
	\label{eqn:rateFemtoCom}
	R_{mf,F}^n = \log_2 \Biggl( 1 + \frac{|h_{mf}^{n}|^2\theta P_m^n}{N_0 + I_k^n  } \Biggr),
\end{equation}
where the interference term is due to: \emph{(i)}- the power used to transmit the coarse messages of other MUEs and \emph{(ii)}- the transmissions from other interfering FUEs.

The relayed FBS signal over the wireless backhaul includes the fine messages of both the MUEs and the FUEs in which a rate fraction of $\nu R_{f0}^n$ is allocated for the MUE's fine message and $(1-\nu) R_{f0}^n$ to the FUE's signal, where $0 <\nu \leq 1$. Since the uplink rate of the backhaul is interference-limited, the rate of the relayed signal using DF relaying is  the minimum rate of the MUE-FBS link and  FBS-MBS backhaul.
Therefore, the total throughput of the $m$-th MUE's {\it fine} message is:
\begin{equation}\label{eqn:relayedrate_OTA}
	[R_{m,F}^n]_{\rm OTA} = \frac{1}{2} \min\{R_{mf,F}^n,\nu R_{f0}^n\}.
\end{equation}
The factor $\frac{1}{2}$ accounts for the half duplex DF relaying constraint and the total MUE rate  is the sum of (\ref{eqn:rateMacroPriv}) and (\ref{eqn:relayedrate_OTA}).

\subsection{Wired backhaul (WRD)}
In the wired backhaul scenario, the backhaul capacity which is constrained by (\ref{eqn:totwiredcapacity}) influences the final rate of the relayed fine message. Moreover, the rates of the coarse and fine messages of MUE $m$ are given by:
\begin{align}
\label{eqn:rateMacroPriv_wire}
	[R_{m,C}^{n}]_{\rm WRD} = \log_2 \Biggl( 1 + \frac{|h_{m0}^{n}|^2(1-\theta)P_m^n}{N_0 +  \sum_{\forall j\in\mathcal{K}^n}|h_{j0}^{n}|^2P_{j}^n } \Biggr)
\end{align}
\begin{equation}
\label{eqn:relayedrate_wire}
	\textstyle [R_{m,F}^n]_{\rm WRD} = \frac{1}{2}\min\Biggl\{ \log_2 \biggl( 1 + \frac{|h_{mf}^{n}|^2\theta P_m^n}{N_0 + I_k^n - |h_{mf}^{n}|^2\theta P_m^n } \biggl), \nu C_f \Biggl\},
\end{equation}
where the fraction of the capacity allocated for the MUE's fine messages is $\nu C_f$.

\section{Game-Theoretic Approach for Enabling Cooperative Relaying}\label{sec:game}
\subsection{Game Formulation}
In order to benefit from rate splitting and femtocell relaying, the MUEs must be able to appropriately choose their preferred FBS, given the channel conditions as well as the underlying backhaul constraints. In this regard, we formulate a non-cooperative game $\mathcal{G}= \Big(\mathcal{M},\{\mathcal{A}_m\}_{m \in \mathcal{M}},U(a_m,\boldsymbol{a}_{-m})\Big)$ in which $\mathcal{M}$ denotes the set of players, i.e., the MUEs, $\mathcal{A}_m$ is the action set taken by MUEs which represents the discrete power level ($\theta$ and $P_m^n$) and the chosen relaying FBS from the set $\mathcal{F}$, and $U(a_m,\boldsymbol{a}_{-m})$ is the utility function of each MUE $m$ with an action $a_m\in\mathcal{A}_m$ while $\boldsymbol{a}_{-m}$ denotes the vector of actions taken by all other MUEs (players) except $m$.
The utility should capture the tradeoff between the achieved  throughput from relaying and the expected delay due to the backhaul constraints. One suitable metric for capturing the tradeoff between throughpout and delay is that of a  \emph{system power } which is defined as the ratio of some power of the throughput and the delay~\cite{PW01,pap:vladimir06}. Hence, using this metric, the utility  function of any MUE $m$ that chooses action $a_m\in\mathcal{A}_m$ can be defined as:
\begin{equation}
	U(a_m,\boldsymbol{a}_{-m}) = \frac{[R_m(a_m,\boldsymbol{a}_{-m})]^\delta}{[D_m(a_m,\boldsymbol{a}_{-m})]^{(1-\delta)}},
\end{equation}
where $R_m(a_m,\boldsymbol{a}_{-m})$ is the total rate and $D_m(a_m,\boldsymbol{a}_{-m})$ is the total delay experienced by MUE $m$ based on actions $(a_m,\boldsymbol{a}_{-m})$ which map to the individual $\theta$ and $P_i^n,~\forall i\in\mathcal{M}$, and $\delta\in(0,1)$ is a parameter that highlights the sensitivity of the MUE's service to throughput and delay~\cite{PW01,pap:vladimir06}. 

We note that the choice of an action $a_m \in \mathcal{A}_m$ by an MUE $m$ implies the use of a couple $(\theta,P_m^n)=(\theta',{P_m^n}')$. Subsequently, the rate $R_m(a_m,\boldsymbol{a}_{-m})$ can be found using (\ref{eqn:rateMacroPriv})-(\ref{eqn:relayedrate_wire}) depending on the backhaul (wired or wireless), with $\theta'$ and ${P_m^n}'$ representing the action $a_m$ selected by MUE $m$. Similarly, the delay $D_m(a_m,\boldsymbol{a}_{-m})$ is
dependent on the action $a_m$ and this dependence is shown later in this section in (\ref{eqn:delayClasic})-(\ref{eqn:delayProposed2}).
For notational simplicity, hereinafter, we use $R_m$ and $D_m$ to denote, respectively, $R_m(a_m,\boldsymbol{a}_{-m})$ and $D_m(a_m,\boldsymbol{a}_{-m})$.

In the classical approach, the transmission delay for MUE is expressed by \cite[pp.741]{book:gracia08};
\begin{equation}\label{eqn:delayClasic}
	[D_m^n]_{CLA} = \frac{\lambda_m}{2[R_m^n]_{CLA}([R_m^n]_{CLA}-\lambda_m)},
\end{equation}
where the rate of MUE is given by (\ref{eqn:noncop_mbsrate}).

In the proposed cooperative approach, when rate splitting is used, both coarse and fine messages experience different rates and delays. Indeed,  the coarse message is dependent on the direct link between MUE and MBS, whereas the fine message depends on both MUE-FBS and FBS-MBS links. Both delays are given by:
\begin{align}\label{eqn:delayProposed}
	&[D_{m,C}^n]_{Z} = \frac{\lambda_{m,C}}{2[R_{m,C}^n]_{Z}([R_{m,C}^n]_{Z}-\lambda_{m,C})},\\
\label{eqn:delayProposed2}
	& [D_{m,F}^n]_{Z} = \biggl( \underbrace{\textstyle \frac{\lambda_{m,F}}{2R_1(R_1-\lambda_{m,F})}}_\text{MUE-FBS delay} + \underbrace{\textstyle \frac{\lambda_{m,F}}{2R_2(R_2-\lambda_{m,F})}}_\text{MBS-FBS delay} \biggr),
\end{align}
where the subscript $Z$ refers to either the wired or the wireless backaul case, and $\lambda_{m,C},\lambda_{m,F}$ represent the packet generation rates for the coarse and fine messages, respectively. Irrespective of the backhaul type, we have $R_1=R_{mf,F}^n$. In addition, in the wired scenario, $R_2$ is given by $\nu C$ while in the OTA case $R_2$ is given by $\nu R_{f0}^n$.
Note that, although the MUEs generate packets for both the coarse and the fine messages at the same instance, due to their channel conditions, the corresponding delays are essentially different  as per (\ref{eqn:delayProposed}). In order to complete the entire transmission, all packets should reach the MBS and therefore the entire transmission time  is the largest delay out of $[D_{m,C}^n]_{Z}$ and $[D_{m,F}^n]_{Z}$. Thus, the total delay is the maximum between the delays of the coarse and fine messages. Finally, the utilities of MUE $m$ for both  wired and OTA schemes are:
\begin{equation}
\label{utility}
	[U(a_m,\boldsymbol{a}_{-m})]_{Z} = \frac{([R_{m,C}^n]_{Z}+[R_{m,F}^n]_{Z})^\delta}{(\max\{[D_{m,C}^n]_{Z},[D_{m,F}^n]_{Z}\})^{(1-\delta)}}.
\end{equation}

\subsection{Game Solution and Proposed Algorithm}
In the proposed non-cooperative game, the choices of the MUEs are discrete and relate essentially to the choice of a serving FBS and the associated power level. This type of games is reminiscent of the framework of network formation games~\cite{NET01} in which individuals interact in order to decide on the friendship relationships or links that they wish to form. The solution of a network formation game is essentially a Nash network~\cite{NET01}, which is a Nash equilibrium of the game that constitutes a stable network in which individuals are interconnected through a graph with each link having an  associated ``strength'' or intensity. Similarly, for the proposed MUEs game, the sought solution is essentially a stable Nash network in which no MUE can improve its utility by unilaterally changing its chosen FBS nor the associated power level. Finding analytical closed-form solutions on the existence and properties of a network formation game's equilibrium is known to be a challenging task, notably under generic utility functions such as the one proposed in this work~\cite{NET01}.

\begin{table}[!t]
  \centering
  \caption{
    \vspace*{-0.0em}Proposed network formation algorithm.}\vspace*{-1em}
    \begin{tabular}{p{8cm}}
      \hline
      \textbf{Initial State} \vspace*{.3em} \\
      \hspace*{1em}The starting system consists of MUEs that are directly connected \vspace*{0.01cm}\\
\hspace*{1em}to the MBS.\vspace*{0.01cm}\\
\textbf{The proposed algorithm consists of three phases} \vspace*{.3em}\\
\hspace*{1em}\emph{Phase~I - Neighboring Femtocell Discovery:}   \vspace*{.1em}\\
\hspace*{2em}\textbf{repeat}\vspace*{.2em}\\
\hspace*{3em}Each MUE $m\in\mathcal{M}$ monitors the RSSI of FBSs over the pilot\vspace*{.1em}\\
\hspace*{3em}channel.\vspace*{.2em}\\
\hspace*{3em}Each MUE chooses FBS $f\in\mathcal{F}$ having the largest RSSI which\vspace*{.1em}\\
\hspace*{3em}has not been selected by any other MUE previously.\vspace*{.2em}\\
\hspace*{2em}\textbf{until} all MUEs discover their respective relaying FBSs.\vspace*{.2em}\\

\hspace*{1em}\emph{Phase~II - Iterative Network Formation Algorithm:} 
\vspace*{.1em}\\
\hspace*{2em}\textbf{repeat}\vspace*{.2em}\\
\hspace*{3em}In a random but sequential order, all MUEs form the network.\vspace*{.2em}\\
\hspace*{3em}In every iteration $t$, each MUE $m$ plays its best response by\vspace*{.1em}\\
\hspace*{3em}selecting $\theta$ and $P_m^n$ that maximizes $U(a_m,a_{-m})$.\vspace*{.2em}\\
\hspace*{2em}\textbf{until} convergence to a final Nash solution after $T$ iterations.\vspace*{.2em}\\

\hspace*{1em}\emph{Phase~III - Rate Splitting:}   \vspace*{.1em}\\
\hspace*{3em}Based on the chosen actions, the MUEs perform the proposed\vspace*{.1em}\\
\hspace*{3em}rate-splitting technique.\vspace*{.2em}\\
\hspace*{3em}Supporting FBSs simulataneously decode the MUE's fine mes-\vspace*{.1em}\\
\hspace*{3em}sage and their own FUEs messages using SIC.\vspace*{.2em}\\
\hspace*{3em}Each FBS transmits both FUE and fine signals over the\vspace*{.1em}\\
\hspace*{3em}network's backhaul.\vspace*{.2em}\\
   \hline
    \end{tabular}\label{tab:algo}
\end{table}

However, to overcome this complexity, one can develop algorithmic approaches that can be adopted by the MUEs so as to reach the equilibrium of this game. In this respect, we propose a myopic algorithm, based on best response dynamics~\cite{NET01,tech:derks08}, using which the MUEs interact, iteratively, so as to choose their preferred relaying FBS. 
The proposed algorithm is composed of three main steps:
\begin{enumerate}
\item \textbf{Step~1: Neighboring Femtocell Discovery:} Each MUE discovers prospective relaying FBSs by monitoring the received signal strength indicator (RSSI) over pilot channels.
\item \textbf{Step~2: Iterative Network Formation Algorithm:} Following the discovery phase, each MUE chooses its best response which consists of optimizing its \emph{current} utility by choosing the relaying FBS and the associated power level. Step~2 is repeated until convergence to the Nash solution of the game.
\item \textbf{Step~3: Rate Splitting:} Once the final network forms, the MUEs can perform the proposed rate-splitting technique in which the helping FBS simultaneously decodes the MUE's fine message  and its own FUE using SIC. Subsequently, the helping FBS transmits both MUE and FUE signals over the heterogeneous backhaul.
\end{enumerate}

\section{Simulation Results}\label{sec:results}
For simulations, we consider a single macrocell with radius $R_m=400$ m in which a number of MUEs and FBSs with radius $R_f=20$~m are deployed. Each FBS serves a single FUE. The maximum transmission power is set to $20$ dBm and the noise level is set to -$130$ dBm. We use the 3GPP specifications for path loss and shadowing in both indoor and outdoor links~\cite{FemtoBook}. The shadowing standard deviation is set to $10$~dB while the wall penetration loss is set to $12$~dB. Moreover, both MUEs and FUEs have packet generation process with rates  $\lambda_m=\lambda_f=150$ Kbps, respectively. A super-frame is assumed to be a bundle of $10$ frames (packets)~\cite{book:Boudour08}.

\begin{figure}[!t]
\centering
\subfigure[ Utility per MUE as a function of the number of FBSs ($M=5$, $37.5$ Mbps wired backhaul and $32$ OTA backhaul channels).]{
\includegraphics[width=8cm]{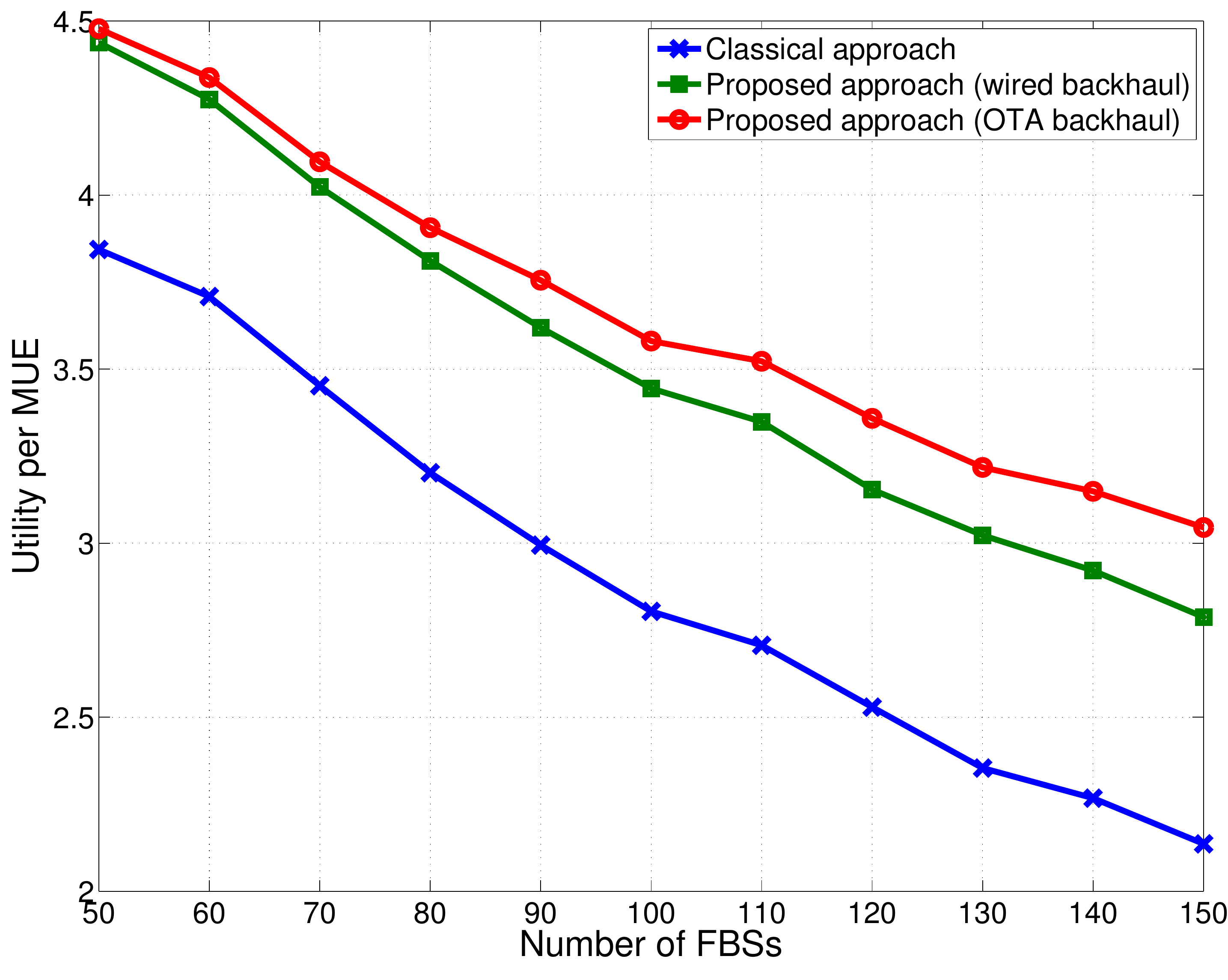}
\label{fig:utFBSchangeOri}
}
\subfigure[Average utility per MUE as a function of the number of MUEs ($F=40, 80$, $37.5$ Mbps wired backhaul and $32$ OTA backhaul channels).]{
\includegraphics[width=8cm]{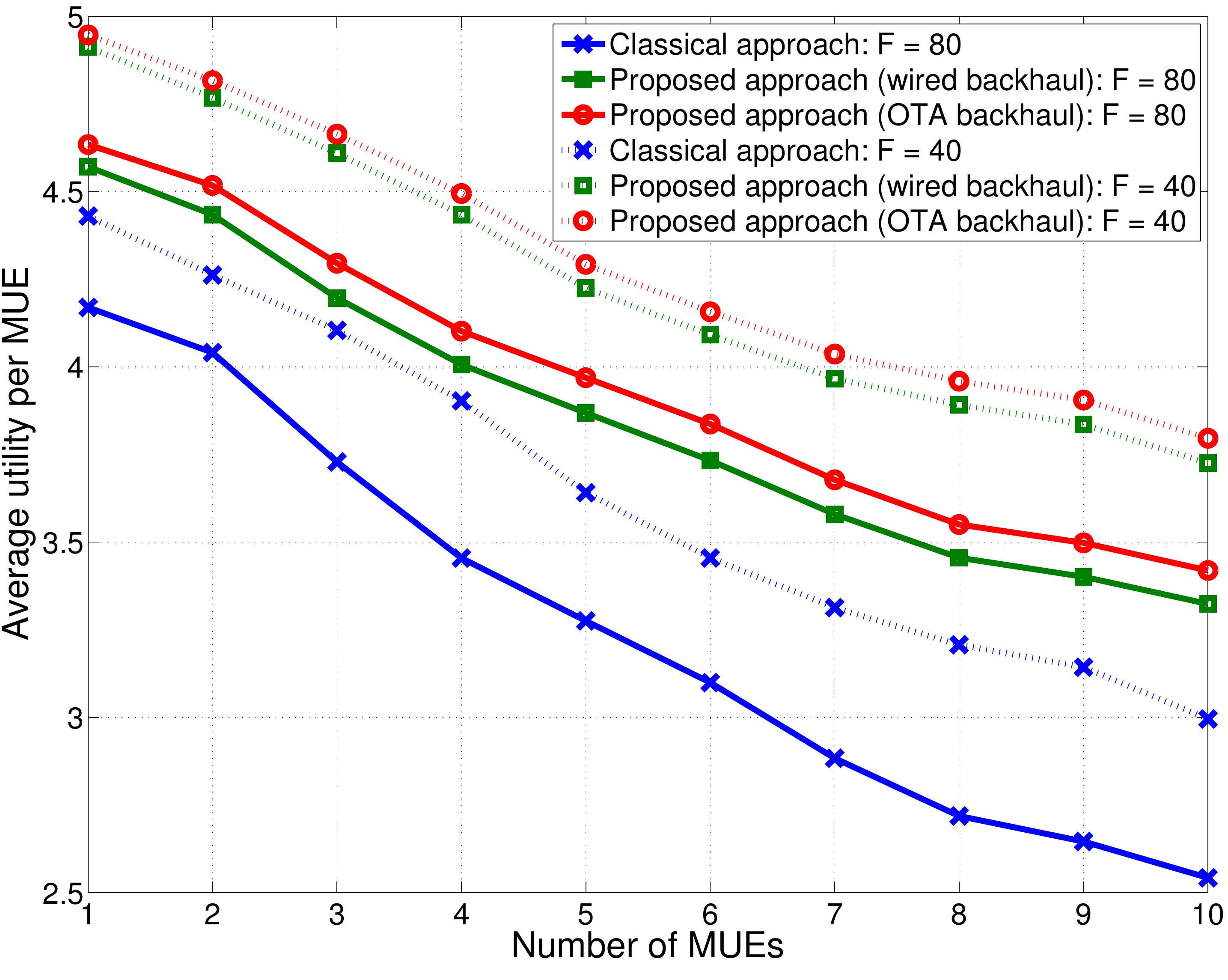}
\label{fig:utMUEchangeOri}
}
\caption[Optional caption for list of figures]{The variations of the average utility per MUE resulting from classical approach and the proposed scheme (fr both wired and wireless backhauls) as a function of the network size.}
\label{fig:utilityChangeori}
\end{figure}

Fig.~\ref{fig:utFBSchangeOri} and Fig.~\ref{fig:utMUEchangeOri} show the average MUE payoff  as a function of the number of FBSs and MUEs, respectively.  These figures show that the proposed approach yields a significant improvement over the classical approach reaching up to $140\%$ for  $F=150$ femtocells.  Furthermore, Fig.~\ref{fig:utFBSchangeOri} and Fig.~\ref{fig:utMUEchangeOri} show that, as the network becomes denser, the average MUE payoff decreases due to the lower rates, increased interference, and higher delays. Nonetheless, for dense networks, it is more likely to have MUEs and FBSs close to one another which  enables the MUEs to achieve a higher payoff using the proposed approach as opposed to the classical method with no coordination. In addition, we can observe that the gaps between proposed scheme and classical case are increasing as the system gets denser, i.e., the drop of utility is low for the proposed schemes compared to the classical scheme. 

\begin{figure}[!t]\label{fig:CDFs}
\centering
\subfigure[CDF of transmission delays.]{
\includegraphics[width=8cm]{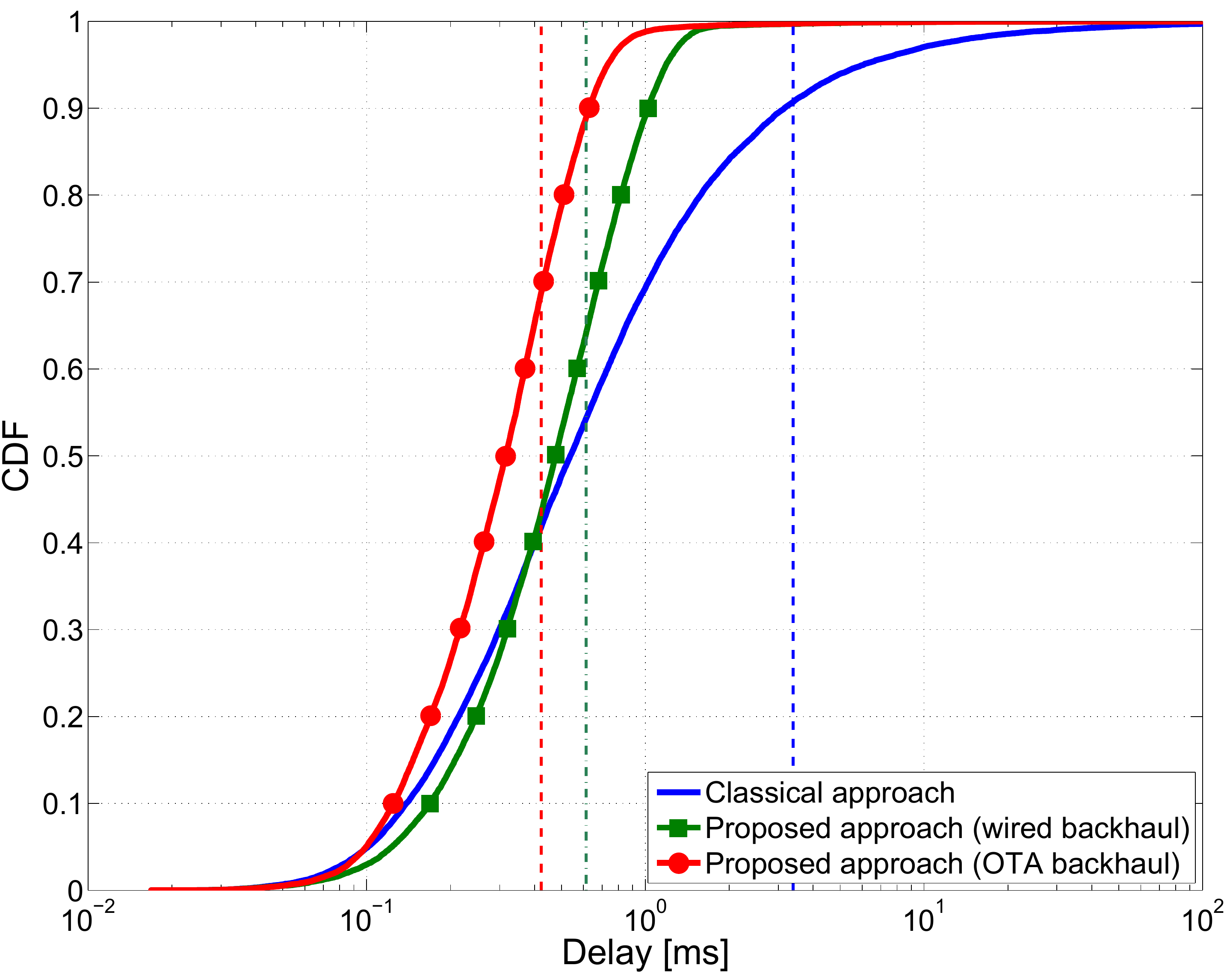}
\label{fig:delayCDF}
}
\subfigure[CDF of transmission rates.]{
\includegraphics[width=8cm]{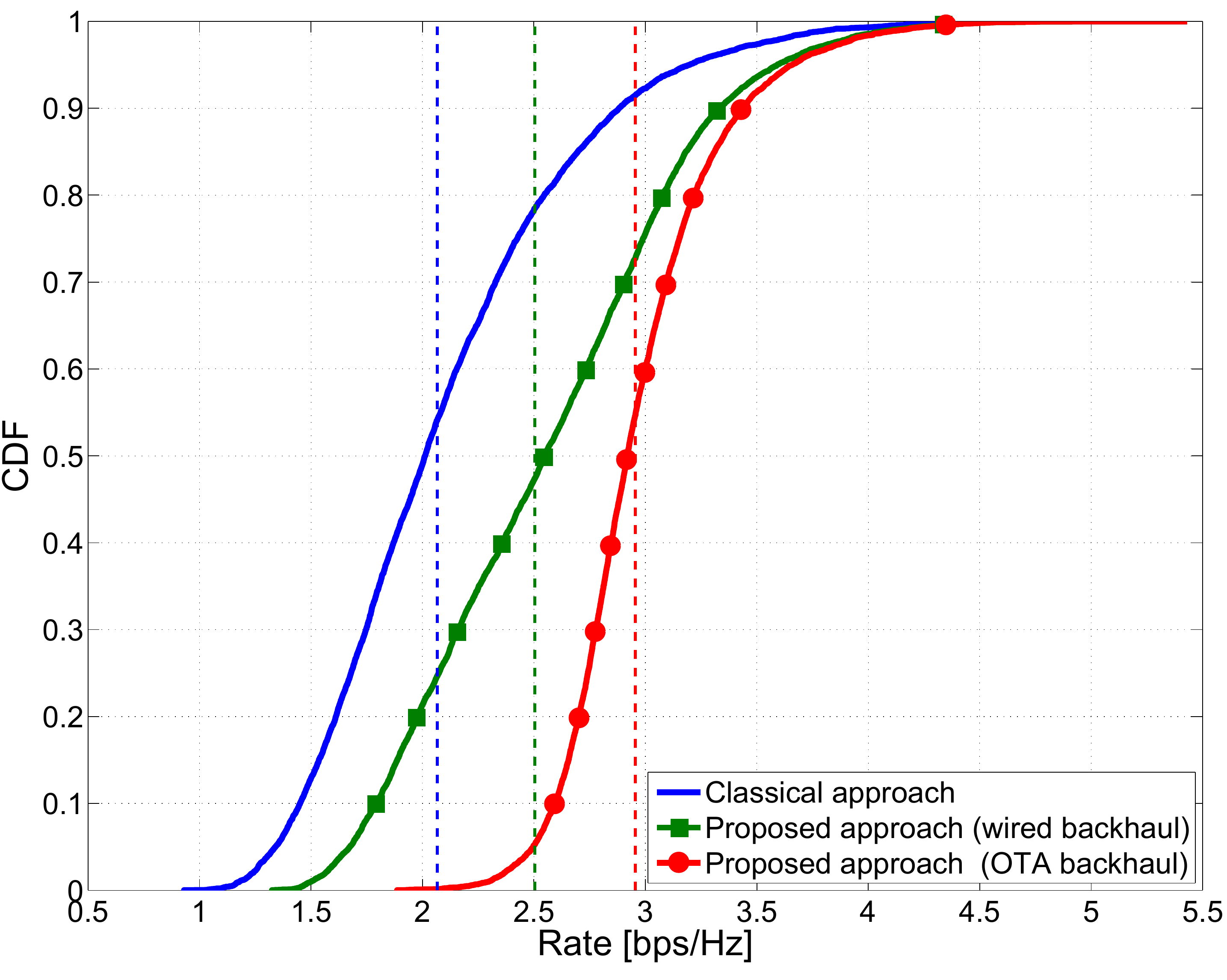}
\label{fig:rateCDF}
}
\caption{Cumulative density functions (CDFs) for the 3 schemes; classical, proposed wired and OTA where the mean values are indicated by vertical dotted lines ($M=\{1,10\}$, $F=\{50,60,\ldots,150\}$, 37.5 Mbps wired backhaul and 32 OTA backhaul channels).}
\end{figure}

The utility figures describe the combined behavior of rates and delays. However, users have interest on improve both rates and delays separately, it is interesting to observe the rate and the delay separately. therefore, the cumulative density functions (CDFs) of transmission delays and rates are presented in Fig.~\ref{fig:delayCDF} and Fig.~\ref{fig:rateCDF}, respectively.

In these two figures, we  observe a reduction in the delays and an improvement in the data rates when using the proposed approaches (wired and OTA) as compared to the classical approach. This is due to the fact that the relaying path provides additional rate gains due to the higher capacity of the MUE-FBS link and the FBS-MBS backhaul (as opposed to direct transmission). In fact, Fig.~\ref{fig:delayCDF} shows that the proposed approach can reduce the delay by $5$ times for the wired backhaul case and by $10$ times for the OTA case. For transmission rates, both proposed wired and OTA schemes improve the average rate of classical approach by  $125\%$ and $150\%$, respectively as shown in Fig.~\ref{fig:rateCDF}.

\begin{figure}[!t]
\centering
\includegraphics[width=8cm]{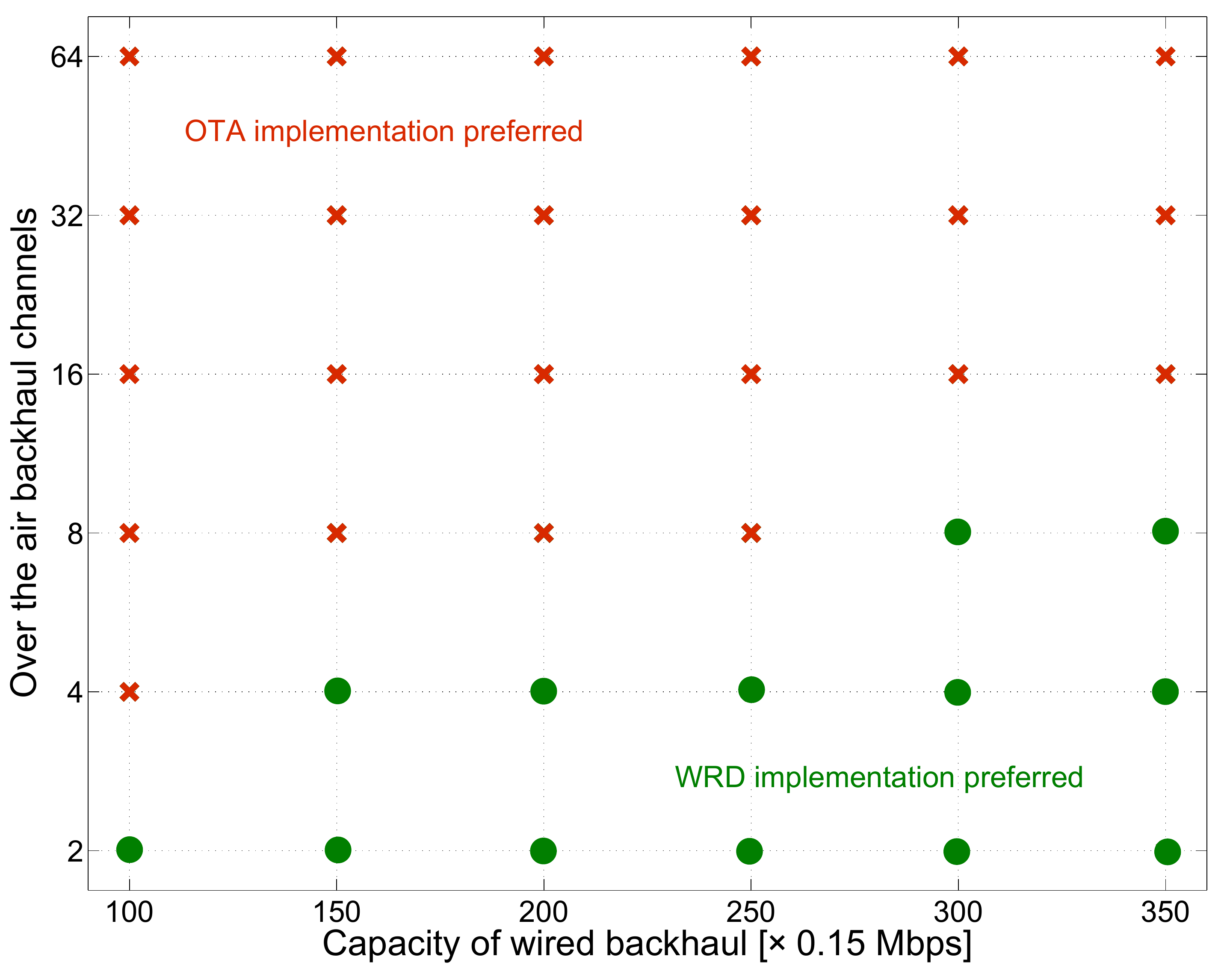}
\caption{The impact of the backhaul capacity (wired) and number of channels (wireless) on the average achievable utility per MUE. The backhaul implementation which provides the best average utility is indicated at each point/combination (the above simulation is performed for a system with $M=5$ and $F=80$).}
\label{fig:WRDOTAtradeoff}
\end{figure}

Fig.~\ref{fig:WRDOTAtradeoff} compares the achievable average utility per MUE for a system with $M=5$ MUEs and $F=80$ FBSs under different backhaul constraints. 
As the capacity of wired backhaul and the number of channels in the wireless backhaul are increased, the average MUE utility is increased, in both cases. 
However, based on the capacity, a certain backhaul implementation (wired or OTA) may provide a higher average utility over the other. As an example, the wired backhaul with 45 Mbps capacity offers a higher average utility compared to the wireless backhual with 8 (or less) channels while it is less compared to an OTA backhaul with 16 (or greater) channels. This tradeoff can be used to define the appropriate backhaul mechanism (wired or OTA) based on the resource availability for each.

\section{Conclusion}\label{sec:conclu}
In this paper, we have proposed rate-splitting techniques so as to enhance the performace of MUEs with aid of open-access femtocells over a heterogeneous backhaul. The performance of MUEs is evaluated by a metric - system power - which captures both achieved throughput and the expected backhaul delay. Simulation results have shown that the proposed approach yields $50\%$ of improvement in throughput and $10$ times reduction in expected delays, compared to the system with no cross-tier cooperation is available. 
Furthermore, we have shown that the proposed approach can capture the tradeoff between two backhaul implementation techniques -- wired and over-the-air - based on the achievable average utility. The proposed scheme can thus be used to determine the appropriate backhaul implementation whenever the knowledge of resource availability for both wired and over-the-air backhauls is known a priori. 
\section*{Acknowledgment}

The authors would like to thank the Finnish funding agency for technology and innovation, Elektrobit, Nokia and Nokia and Siemens Networks for supporting this work. This work has been performed in the framework of the ICT-4-248523 BeFEMTO, which is funded by the EU and the LOCON Project, TEKES, Finland and the Acedemy of Finland grant 128010.

\def\baselinestretch{0.8}
\bibliographystyle{IEEEtran}
\bibliography{IEEEabrv}

\begin{thebibliography}{10}
\providecommand{\url}[1]{#1}
\csname url@samestyle\endcsname
\providecommand{\newblock}{\relax}
\providecommand{\bibinfo}[2]{#2}
\providecommand{\BIBentrySTDinterwordspacing}{\spaceskip=0pt\relax}
\providecommand{\BIBentryALTinterwordstretchfactor}{4}
\providecommand{\BIBentryALTinterwordspacing}{\spaceskip=\fontdimen2\font plus
\BIBentryALTinterwordstretchfactor\fontdimen3\font minus
  \fontdimen4\font\relax}
\providecommand{\BIBforeignlanguage}[2]{{%
\expandafter\ifx\csname l@#1\endcsname\relax
\typeout{** WARNING: IEEEtran.bst: No hyphenation pattern has been}%
\typeout{** loaded for the language `#1'. Using the pattern for}%
\typeout{** the default language instead.}%
\else
\language=\csname l@#1\endcsname
\fi
#2}}
\providecommand{\BIBdecl}{\relax}
\BIBdecl

\bibitem{pap:vikram08}
V.~Chandrasekhar, J.~G. Andrews, and A.~Gathere, ``Femtocell networks: A
  survey,'' \emph{{IEEE} Commun. Mag.}, vol.~46, no.~9, pp. 59--67, Sep. 2008.

\bibitem{pap:jeff12}
J.~G. Andrews, H.~Claussen, M.~Dohler, S.~Rangan, and M.~C. Reed, ``Femtocells:
  Past, present, and future,'' \emph{{IEEE} J. Sel. Areas Commun.}, Apr. 2012.

\bibitem{onln:eric11}
\BIBentryALTinterwordspacing
U.~Ewaldsson and C.~Hedelin, ``Ericsson hetnet,'' Tech. Rep., 2011. [Online].
  Available:
  \url{http://www.slideshare.net/fullscreen/zahidtg/ericsson-hetnet/1}
\BIBentrySTDinterwordspacing

\bibitem{pap:ping11}
P.~Xia, H.-S. Jo, and J.~G. Andrews, ``Fundamentals of intra-cell overhead
  signalling in heterogeneous cellular networks,'' \emph{Computing Research
  Repository (CoRR)}, vol. abs/1106.5825, pp. 1--26, 2011.

\bibitem{pap:johnp10}
J.~P.~M. Torregoza, R.~Enkhbat, and W.-J. Hwang, ``Joint power control, base
  station assignment, and channel assignment in cognitive femtocell networks,''
  \emph{EURASIP Journal on Wireless Communications and Networking}, pp. 1--14,
  2010.

\bibitem{pap:ivana11}
I.~Mari\'c, B.~Bo\^stjan\^ci\^c, and A.~Goldsmith, ``Resource allocation for
  constrained backhaul in picocell networks,'' \emph{Information Theory and
  Applications Workshop (ITA)}, pp. 1--6, Feb. 2011.

\bibitem{pap:osvaldo10}
O.~Simeone, E.~Erkip, and S.~S. Shitz, ``Robust transmission and interference
  management for femtocells with unreliable network access,'' \emph{{IEEE} J.
  Sel. Areas Commun.}, vol.~28, no.~9, pp. 1469--1478, Dec. 2010.

\bibitem{book:dim92}
D.~P. Bertsekas and R.~Gallager, \emph{Data Networks}.\hskip 1em plus 0.5em
  minus 0.4em\relax Prentice Hall International, Inc., 1992.

\bibitem{pap:randa11}
R.~Zakhour and D.~Gesbert, ``Optimized data sharing in multicell {MIMO} with
  finite backhaul capacity,'' \emph{{IEEE} Trans. Signal Process.}, vol.~PP,
  no.~99, pp. 1--10, Aug. 2011.

\bibitem{pap:osvaldo09}
O.~Simeone, O.~Somekh, E.~Erkip, H.~V. Poor, and S.~S. Shitz, ``A broadcast
  approach to robust communications over unreliable multi-relay networks,''
  \emph{Information Theory and Applications Workshop}, pp. 334--340, Feb. 2009.

\bibitem{book:tse05}
D.~Tse and P.~Viswanath, \emph{Fundamentals of Wireless Communication}.\hskip
  1em plus 0.5em minus 0.4em\relax Cambridge University Press, 2005.

\bibitem{PW01}
E.~Altman, T.~Ba\c{s}ar, and R.~Srikant, ``Nash equilibria for combined flow
  control and routing in networks: asymptotic behaviour for a large number of
  users,'' \emph{{IEEE} Trans. Automtatic Control}, vol.~47, pp. 917--930, jun
  2002.

\bibitem{pap:vladimir06}
V.~Vukadinovi\'c and G.~Karlsson, ``Video streaming in 3.5{G}: On
  throughput-delay performance of proportional fair scheduling,'' \emph{IEEE
  International Symposium on Modeling, Analysis, and Simulation of Computer and
  Telecommunication Systems (MASCOTS) 2006}, pp. 393 -- 400, Sep. 2006.

\bibitem{book:gracia08}
A.~Leon-Garcia, \emph{Probability, Statistics, and Random Processes for
  Electrical Engineering}.\hskip 1em plus 0.5em minus 0.4em\relax Pearson
  Prentice Hall, 2008.

\bibitem{NET01}
M.~O. Jackson, \emph{Social and Economic Networks}.\hskip 1em plus 0.5em minus
  0.4em\relax Princeton, NJ, USA: Princeton University Press, Dec. 2010.

\bibitem{tech:derks08}
J.~Derks, J.~Kuipers, M.~Tennekes, and F.~Thuijsman, ``Local dynamics in
  network formation,'' Maastricht Universitiy, Department of Mathematics, Tech.
  Rep., Oct. 2008.

\bibitem{FemtoBook}
J.~Zhang and G.~de~la Roche, \emph{Femtocells: Technologies and
  Deployment}.\hskip 1em plus 0.5em minus 0.4em\relax West Sussex, UK: John
  Wiley \& Sons, Ltd., Mar. 2010.

\bibitem{book:Boudour08}
G.~Boudour, C.~Teyssi\'e, and Z.~Mammeri, \emph{Wireless and Mobile
  Networking}.\hskip 1em plus 0.5em minus 0.4em\relax SpringerLink, 2008, vol.
  284/2008, ch. Performance Analysis of Reservation {MAC} Protocols for Ad-hoc
  Networks.

\end{thebibliography}

\end{document}